\documentclass[prd,aps,twocolumn,showpacs,preprintnumbers,amsmath,amssymb]{revtex4} 
\usepackage{graphicx}
\usepackage[]{caption} 
\usepackage{amsmath} 
\usepackage{amssymb} 
\def\beq{\begin{equation}}
\def\eeq{\end{equation}}
\def\bea{\begin{eqnarray}}
\def\eea{\end{eqnarray}}

\begin{document} 
 
\preprint{SU-4252-877} 
 
\title{Hamilton-Jacobi Formalism for String Gas Thermodynamics} 

\author{Anosh Joseph$^{a}$}\thanks{ajoseph@phy.syr.edu} \author{Sarada G. Rajeev$^{b}$}\thanks{rajeev@pas.rochester.edu}
\affiliation{$^{a}$Department of Physics, Syracuse University, Syracuse, NY
13244-1130, USA \\
$^{b}$~Department of Physics and Astronomy and Department of Mathematics, University of Rochester, Rochester, NY 14627-0171, USA}
\date{\today} 
 
\begin{abstract} 

We show that the thermodynamics of a system of strings at high energy densities under the ideal gas approximation has a formulation in terms of Hamilton-Jacobi theory. The two parameters of the system, which have dimensions of energy density and number density, respectively, define a family of hypersurfaces of co-dimension one, which can be described by the vanishing of a function $F$ that plays the role of a Hamiltonian.

\end{abstract} 
 
\maketitle 
\section{INTRODUCTION}
The similarity between the structure of classical mechanics and geometric optics was known to the nineteenth century mathematicians W. Hamilton and C. Jacobi. The eikonal equation in geometric optics and the Hamilton-Jacobi equation in classical mechanics serve the same purpose: they are both approximations to wave theories in the limit of short wavelength. Classical thermodynamics too has a geometrical meaning. It was first realized by J. Gibbs \cite{Gibbs} when he combined the first and second laws of thermodynamics and later it was developed using the theory of differential forms by J. Pfaff, C. Carath\'eodory and others. (See Ref. \cite{Arnold} for a very detailed description of the geometrical formulation.) 

The physical variables in classical thermodynamics, like in classical mechanics, occur in conjugate pairs. But the phase space of classical thermodynamics, unlike the phase space of classical mechanics, is odd dimensional. It is a {\it contact manifold}. Classical thermodynamics can be formulated in terms of contact geometry \cite{Rajeev1, Rajeev2}. A substance which has $n$ degrees of freedom lives in a $2n+1$ dimensional thermodynamic phase space. The one-form with `coordinates' $q^{i}$ and `conjugate momenta' $p_{i}$,
\beq
\label{eq:contact-form}
\alpha \equiv dq^{0} - p_{i}dq^{i},
\eeq 
defines a contact structure on the thermodynamic phase space. 

There are many coordinate systems in which the contact form given in Eq. (\ref{eq:contact-form}) preserves the same structure. Transformations from one coordinate system to another that preserve the above contact form are called {\it Legendre transformations}. Infinitesimal Legendre transformations are determined by a single function $F$ called the {\it generating function} or {\it Hamiltonian}. The function $F$ plays a crucial role in the context of the dynamics of thermodynamics. From the generating function we can have a set of ordinary differential equations that are analogues of Hamilton's equations in classical mechanics
\bea
\frac{dq^{i}}{dt} = \frac{\partial F}{\partial p_{i}},~\frac{dp_{i}}{dt} = -\frac{\partial F}{\partial q^{i}}-p_{i}\frac{dF}{dq^{0}},~\frac{dq^{0}}{dt} = p_{i}\frac{\partial F}{\partial p_{i}}-F.
\eea 
The solutions of these equations are called the {\it characteristic curves} of the generating function $F$. These curves define a dynamical system on every hypersurface
\beq
\label{eq:hyper}
F(q^{0}, q^{1}, \cdots, q^{n}, p_{1}, \cdots, p_{n}) =0,
\eeq
on a contact manifold.  

The hypersurface given in Eq. (\ref{eq:hyper}) defines the dynamics on the thermodynamic phase space. It is the Hamilton-Jacobi equation for the thermodynamic system.

As in the case of a system of point particles, a system of strings can also be described on a thermodynamic phase space. In this paper we discuss the Hamilton-Jacobi formalism of classical thermodynamics for an ideal gas of strings at high energy densities. In \cite{Rajeev2} Hamilton-Jacobi equations are derived for the cases of a system of van der Waals gases, Curie-Weiss magnets and Schwarzschild-Anti de Sitter family of black holes. Recently, there have been much progress in the area of string gas cosmology \cite{Brandenberger1, Bassett, Ali, Borunda, Brandenberger2}, a scanerio of the very early universe based on the new degrees of freedom and new symmetries inspired by string theory (See \cite{Brandenberger3} for a brief review and \cite{Battefeld} for an extensive critical review.), and its most recent incarnation as a Hagedorn bounce \cite{Kaloper:2007pw, Kaloper:2006xw}. Hamilton-Jacobi formalism of classical thermodynamics leads to a route to quantum thermodynamics \cite{Rajeev1, Coutant}. Although quantum thermodynamics is not the main focus of this paper, it would be interesting to study the effects of quantum thermodynamic fluctuations in the hot string gas phase of the very early universe.

A string gas displays many interesting features compared to a system of point particles. They are due to the fact that the number of string degrees of freedom grows exponentially with energy \cite{Hagedorn, Huang-Weinberg}, and there is an organized geometric interaction among strings \cite{Deo1}. At high energies, in the case of strings, the Boltzmann factor $e^{-\beta H}$ exactly compensates the leading linear behavior of entropy, while in the case of point particles, the Boltzmann factor dominates over entropy. 

For the Hamilton-Jacobi formulation, we consider two cases of an ideal string gas based on the topology of the universe in which the string gas lives. In the first case, the string gas lives in a space where at least three spatial dimensions are non-compact and the remaining dimensions are very small (of the order of $\sqrt{\alpha'}$, where $\alpha'$ is the slope parameter) with toroidal geometry. In the second case, the string gas lives in a compactified universe, where at least three of the compactified dimensions are large and expanding. We discuss the Hamilton-Jacobi equation for the two cases. The second case is more interesting in the context of the cosmology of very early universe, where the size of the universe was very small and energy density was very high.
\section{A Primer of past work}
In this section, we summarize the pertinent aspects of the previous work on Hamilton-Jacobi formalism for thermodynamics \cite{Rajeev1, Rajeev2}.

The mathematical structure that captures the essence of thermodynamics must necessarily be odd dimensional. On a manifold with dimension $2n+1$, a {\it contact structure} in a coordinate patch ($q^{0}, q^{1}, \cdots, q^{n}, p_{1}, \cdots, p_{n}$) is given by the one-form
\beq
\label{eq:contact-two}
\alpha \equiv dq^{0} - \sum_{i=1}^{n}p_{i}d q^{i}.
\eeq
The vanishing of the infinitesimal variations in this one-form defines a contact structure. This means that even if we multiply $\alpha$ by a non-zero function $f$, the contact structure is preserved. A contact manifold is a union of coordinate patches. Though all the coordinate patches give equivalent descriptions of the same system, one coordinate patch might give a more simple description than the others. The transformations among coordinate patches can also be defined. They are called Legendre transformations \cite{Rajeev2}. 

An infinitesimal Legendre transformation with a small parameter $\varepsilon$ 
\beq
q^{0} \rightarrow q^{0} + \varepsilon V^{0},~~q^{i} \rightarrow q^{i} + \varepsilon V^{i},~~p_{i} \rightarrow p_{i} + \varepsilon V_{i}
\eeq
defines a vector field
\beq
V = V_{0} \frac{\partial}{\partial q^{0}}+V^{i} \frac{\partial}{\partial q^{i}}+V_{i} \frac{\partial}{\partial p_{i}}
\eeq
whose components can be expressed in terms of a single function 
\beq
\label{eq:F}
F = p_{j}V^{j} - V^{0}.
\eeq

The components of the vector field $V$ are
\beq
V^{i} = \frac{\partial F}{\partial p_{j}},~~V_{i} = -\Big[\frac{\partial F}{\partial q^{i}}+p_{i}\frac{\partial F}{\partial q^{0}}\Big],~~V^{0} = p_{i}\frac{\partial F}{\partial p_{i}} - F.
\eeq
The function $F$ that determines infinitesimal Legendre transformations is called its generating function. It can be shown that the infinitesimal transformation of the contact form given in Eq. (\ref{eq:contact-two}) is unchanged under this vector field \cite{Rajeev2}. 

A thermodynamic substance which has $n$ degrees of freedom lives in a $2n+1$ dimensional thermodynamic phase space. The first law of thermodynamics describes a contact structure on the thermodynamic phase space 
\beq
\alpha \equiv dq^{0} - \sum_{i=1}^{n}p_{i}dq^{i} = 0,
\eeq 
where the `coordinates' $q^{i}$ and `conjugate momenta' $p_{i}$ can be thought of as extensive and intensive variables respectively. (This is not always the case, a Legendre transformation can mix this up \cite{Rajeev2}.) 

When we impose the condition that {\it any infinitesimal change in the state of the substance must satisfy the first law}, we get equations of state of the substance. Equations of state describe an $n$-dimensional sub-manifold called the {\it Lagrangian sub-manifold} and there are $n+1$ of them in a $2n+1$ dimensional phase space. 

Once we have the equation of state called the `fundamental relation' 
\beq
q^{0} = \Phi(q^{1}, \cdots, q^{n}),
\eeq
the remaining $n$ equations of state can be obtained from this by differentiation. For example, in the case of an ideal gas of point particles in three space dimensions, in the picture where entropy $S$ is the thermodynamical potential, we have the fundamental relation $S$ as a function of internal energy $E$ and volume $V$,
\beq
\label{eq:entropy}
S = \textrm{ln}\Big[E^{\frac{3}{2}}~V\Big]
\eeq
up to a constant. The other two equations of state follow from this by differentiation
\bea
\label{eq:temperature}
\frac{1}{T} &=& \Big(\frac{\partial S}{\partial E}\Big)_{V}~,\\
\label{eq:pressure}
\frac{P}{T} &=& \Big(\frac{\partial S}{\partial V}\Big)_{E}~.
\eea
Eqs. (\ref{eq:entropy}), (\ref{eq:temperature}) and (\ref{eq:pressure}) describe the three Lagrangian sub-manifolds for the ideal gas of point particles.

We can think of a family of substances, where each member of the family is specified by the set of parameters, say, ($a_{1}, \cdots, a_{n}$). The simplest example is the family of the van der Waals gases where each member of the family is characterized by the parameters $a_{1}$ (measure of the long range attraction between the gas particles) and $a_{2}$ (measure of the short range repulsion between the gas particles).

For the case of a family of substances the fundamental relation takes the form
\beq
q^{0} = \Phi(q^{1}, \cdots, q^{n}|a_{1}, \cdots, a_{n}).
\eeq
The remaining equations of state follow from the derivatives of the fundamental relation. They are
\beq
p_{i} = \Phi_{i}(q^{1}, \cdots, q^{n}|a_{1}, \cdots, a_{n}),~~~\Phi_{i} = \frac{\partial \Phi}{\partial q^{i}}.
\eeq

The function $\Phi$ defining such a family must satisfy a non-degeneracy condition
\beq
\label{eq:degeneracy}
\textrm{det}\frac{\partial^{2} \Phi}{\partial q~\partial a} \neq 0.
\eeq
Given that the non-degeneracy condition is satisfied, from the $n+1$ equations of state we can get a single function $F (q^{0}, q^{1}, \cdots, q^{n}, p_{1}, \cdots, p_{n})$ relating all the thermodynamic variables. This is the same function $F$ given in Eq. (\ref{eq:F}), in the context of infinitesimal Legendre transformations. 

Once we eliminate the parameters ($a_{1}, \cdots, a_{n}$) from $n+1$ equations of state, we have a single relation
\beq
F(q^{0}, q^{1}, \cdots, q^{n}, p_{1}, \cdots, p_{n}) =0.
\eeq
The function $F (q, p)$ defines a dynamics on the hypersurface of thermodynamic phase space. We encounter a similar function in classical mechanics and there we call it the Hamiltonian or generating function of the mechanical system. Here also we call this function the Hamiltonian of a family of substances.

Thus the thermodynamics of a family of substances with $n$ degrees of freedom is described by a contact manifold of $2n+1$ dimensions with a contact structure $\alpha$ and a family of hypersurfaces $F(q, p) =0$ of co-dimension one on the manifold.

Once the Hamiltonian $F$ of a family of substances is given, the equations of state of each member of the family can be obtained by solving the Hamilton-Jacobi equation
\beq
F\Big(q^{0}, q^{1}, \cdots, q^{n}, \frac{\partial \Phi}{\partial q^{1}}, \cdots,\frac{\partial \Phi}{\partial q^{n}} \Big) = 0.
\eeq 
The choices of the parameters ($a_{1}, a_{2}, \cdots, a_{n}$) will give different equations of state describing each members of the family.

The ordinary differential equations associated with the generating function $F$ 
\bea
\label{eq:curves}
\frac{dq^{i}}{dt} = \frac{\partial F}{\partial p_{i}},\frac{dp_{i}}{dt} = -\frac{\partial F}{\partial q^{i}}-p_{i}\frac{dF}{dq^{0}},\frac{dq^{0}}{dt} = p_{i}\frac{\partial F}{\partial p_{i}}-F~~
\eea 
give solutions that are called the characteristic curves of $F$. These equations are the analogues of Hamilton's equations in classical mechanics.

The parameters ($a_{1}, a_{2}, \cdots, a_{n}$) characterizing a particular member of the family are given by the initial conditions of this dynamics. The complete integral of the Hamilton-Jacobi equation can be obtained once we eliminate the `time' variable $t$ from the above equations. This in turn gives the equations of state that are characterized by the choice of parameters ($a_{1}, a_{2}, \cdots, a_{n}$). It should be noted that the `time' variable of this dynamics is related to entropy \cite{Rajeev2}. 

\section{Hamilton-Jacobi Formalism for String Gas Thermodynamics}
We closely follow the approach of Deo, Jain and Tan \cite{Deo1, Deo2, Deo3, Deo4} for the termodynamics of string gas. The string gas treated under the ideal gas approximation. That is, the interaction between strings is so weak that the spectrum of the theory is the same as that of the free string. The following discussion can be applied to the case of closed bosonic, heterotic and type-II superstrings. (For closed bosonic strings we have to worry about the presence of a tachyon and the infinite cosmological constant. The results are valid for the bosonic case also if the divergences originating from these are ignored.) 

At temperatures close to the Hagedorn temperature the canonical partition function is ill defined and thus it is not useful to deduce the thermodynamic properties of the string gas. Thus one should start from the more fundamental microcanonical ensemble description of statistical mechanics \cite{Frautschi, Carlitz, Mitchell-Turok, Turok}. The system is in equilibrium and it has a fixed total energy $E$. Since there are no conserved quantities other than the energy, {\it an isolated system in equilibrium, with total energy $E$ samples all its eigenstates at that energy with equal probability.} Assuming that this fundamental postulate of statistical mechanics is valid, the microcanonical distribution function or the total density of states $\Omega(E, V)$ gives the consistent definition of the entropy of the system
\beq
S(E, V) \equiv \textrm{ln}~\Omega(E, V).
\eeq  

The first law of thermodynamics says that infinitesimal changes in the thermodynamical quantities, volume $V$, pressure $P$, temperature $T$, entropy $S$ and total energy $E$, must satisfy the contact form
\beq
\alpha \equiv dE-TdS+PdV = 0.
\eeq
It also implies that only two out of the five variables are independent. The remaining variables are given by the equations of state.

When the fundamental relation is known, that is the thermodynamical potential $S$ is given as a function of the coordinates $E$ and $V$, the remaining two equations of state are  
\beq
\frac{1}{T} \equiv \Big(\frac{\partial S}{\partial E}\Big)_{V},~~~\frac{P}{T} \equiv \Big(\frac{\partial S}{\partial V}\Big)_{E}. 
\eeq
\subsection{Ideal Gas of Strings in a Non-compact Universe} 
We assume that at least three Euclidean dimensions are non-compact. The remaining dimensions are very small and compactified on circles of radii $R_{i}, i =1,2, \cdots, D-d$, where $D$ is the total number of spatial dimensions ($D=9$ for heterotic and type-II, $D=25$ for bosonic.) and $d$ is the number of non-compact Euclidean dimensions. Also we assume that the strings can wind around the compact dimensions but not the non-compact ones. We consider the case in which the total energy $E$ is the only conserved quantity in the system. (We do not consider the case where we also have other conserved charges like momenta in the non-compact spatial directions and discrete momenta and windings in the compact directions.)

Assuming that $\epsilon \geq \epsilon_{0}$, where $\epsilon$ is the energy of a single string state and and $\epsilon_{0}$ is the cutoff energy, from  \cite{Deo1, Deo2} we have the microcanonical distribution for the case when $d \neq 0$
\beq
\label{eq:string}
\Omega(E, V) = \frac{CV}{\bar{E}^{\nu}} e^{\beta_{H}E + \lambda_{0}V}\Big[1+O(\frac{1}{\bar{E}})+O(\frac{V}{\bar{E}^{\delta}})\Big],
\eeq
where $\bar{E} \equiv E-\rho_{0}V$, the parameters $\rho_{0}$ and $\lambda_{0}$ have dimensions of energy density and number density respectively, $\delta$ is the smaller of $\frac{d}{2}$ or $2$, $\beta_{H}$ is the inverse Hagedorn temperature \cite{Hagedorn, Huang-Weinberg} given by 
\beq
\beta_{H} = (2 \pi^{2} \alpha')^{1/2}(\sqrt{\omega_{l}}+\sqrt{\omega_{r}}),
\eeq
where $(\omega_{l}, \omega_{r})$ is $(2,2)$, $(2,1)$ and $(1,1)$ respectively for the closed bosonic string, heterotic string  and type-II superstring, $\nu = \frac{d}{2}+1$ and $C=(2\pi^{2} \alpha' \beta_{H})^{-\frac{d}{2}}(\omega_{l}\omega_{r})^{\frac{d}{4}}$. 

It should be noted that $\beta_{H}$ is a universal constant independent of open and closed sectors, the size of the system, and other details of compactification \cite{Antoniadis, Axenides}, but depends only on the type of string theory. The parameters $\rho_{0}$ and $\lambda_{0}$ are dependent on the sector, the cut-off energy $\epsilon_{0}$, and the details of compactification within a given type of string theory, both dimensionally and numerically \cite{Deo1, Abel}. In the thermodynamic limit ($E$ and $V$ large such that $\rho=E/V$ is finite), for the case $d \geq 3$, the correction term in Eq. (\ref{eq:string}) is very small \cite{Deo1, Deo2}. We discuss only the case for which $d \geq 3$.

The entropy of the string gas system is given in terms of the other two extensive variables $E$ and $V$
\beq
\label{eq:S}
S(E, V) = \beta_{H}E + \lambda_{0}V - \nu \textrm{ln}\frac{\bar{E}}{V}
\eeq
up to a constant.

From this fundamental relation the other two equations of state follow immediately
\bea
\label{eq:T}
\frac{1}{T} &=& \beta_{H} - \nu \frac{1}{(\rho - \rho_{0})V},\\
\label{eq:P}
\frac{P}{T} &=& \lambda_{0} + \nu \frac{\rho_{0}}{(\rho - \rho_{0})V} + \frac{1}{V}.
\eea

From Eq. (\ref{eq:T}), it is clear that the temperature $T$ is always just slightly above the Hagedorn temperature $T_{H}$. However it approaches $T_{H}$ in the thermodynamic limit. Eq. (\ref{eq:P}) can be written in the form $PV = \bar{N}T$, where
\beq
\label{eq:N}
\bar{N} = \lambda_{0}V + \nu \frac{\rho_{0}}{\rho - \rho_{0}} + 1.
\eeq
Eq. (\ref{eq:N}) shows that at high energy densities ($\rho \gg \rho_{0}$) the string gas acquires an essentially constant number density $\lambda_{0}$.

We see that the thermodynamic potential $S$ satisfies the non-degeneracy condition,
\beq
\left| \begin{array}{cc}
\frac{\partial^{2} S}{\partial E\partial \rho_{0}}  & \frac{\partial^{2} S}{\partial E\partial \lambda_{0}} \\
\frac{\partial^{2} S}{\partial V\partial \rho_{0}} & \frac{\partial^{2} S}{\partial V\partial \lambda_{0}}  \end{array} \right| = -\frac{\nu V}{(E-\rho_{0}V)^{2}} \neq 0.
\eeq
This non-degeneracy condition indicates the existence of a family of thermodynamic hypersurfaces of co-dimension one. The choices of two parameters $\rho_{0}$ and $\lambda_{0}$ pick out each member of the family.
 
From the three equations of state, Eqs. (\ref{eq:S}), (\ref{eq:T}) and (\ref{eq:P}), the two parameters $\rho_{0}$ and $\lambda_{0}$ can be eliminated to obtain an equation relating the thermodynamic variables:
\beq
\label{eq:hypersurface}
\frac{PV}{T} + \frac{E}{T} - \nu~\textrm{ln}\Big[\frac{\nu}{V} \Big(\frac{1}{T_{H}}-\frac{1}{T}\Big)^{-1}\Big] -S +\nu -1=0.
\eeq
This is the hypersurface $F(S, E, V, \frac{\partial S}{\partial E}, \frac{\partial S}{\partial V})=0$ in the thermodynamic phase space describing the whole family of string gases characterized by the different choices of the parameters ($\rho_{0}$, $\lambda_{0}$) for very high energy densities ($\rho \gg \rho_{0}$). It should be noted that $T_{H}$ in Eq. (\ref{eq:hypersurface}) is a universal constant for a given type of string theory and thus it is the same for all the members of the family given by that hypersurface. 

From the above expression for thermodynamic hypersurface one can obtain the characteristic curves defining the dynamics of the system. It is also interesting to study the form of the variable (or variables) that play the role of `time' in the dynamics of this thermodynamic system.

The ideal gas of strings satisfies the non-degeneracy condition and exhibits a hypersurface connecting all the thermodynamic variables. But for the case of ideal gas of relativistic point particles such a hypersurface, and thus a family of hypersurfaces does not exist. (It is easy to check that from the microcanonical distribution function for an ideal gas of relativistic point particles in a large volume $V$ in $d$ spatial dimensions: $\Omega(E, V) = b V^{-1/2} \rho^{-\gamma} e^{cV\rho^{\alpha}}$, where $\rho \equiv E/V$ is the energy density, $\alpha = d/(d+1)$, $\gamma=(d+2)/(2d+2)$, and $b$, $c$ are dimensionless parameters.) The strange behavior of the ideal string gas can be traced back to the most fundamental property of string theory, namely, its relation to two-dimensional conformal field theory \cite{Deo1, Deo2}. 

The Hamilton-Jacobi equation of the thermodynamics of string gas can be obtained from Eq. (\ref{eq:hypersurface}). For the case in which $S$ is the thermodynamic potential and $E$, $V$ are coordinates, we have the conjugate momenta,
\beq
p_{E} \equiv \Big(\frac{\partial S}{\partial E}\Big)_{V}=\frac{1}{T},~~p_{V} \equiv \Big(\frac{\partial S}{\partial V}\Big)_{E}=\frac{P}{T}. 
\eeq
The Hamilton-Jacobi equation is given by the vanishing of the function $F(S, E, V, p_{E}, p_{V})$:
\beq
\label{eq:hamilton-j}
V p_{V}+ E p_{E}-\nu~\textrm{ln}\Big[\frac{\nu}{V} \Big(\beta_{H}-p_{E}\Big)^{-1}\Big]-S +\nu -1=0.
\eeq
The complete integral \cite{Courant} of this equation will have two parameters, $\rho_{0}$ and $\lambda_{0}$. 
\begin{figure}
\includegraphics[height=8cm]{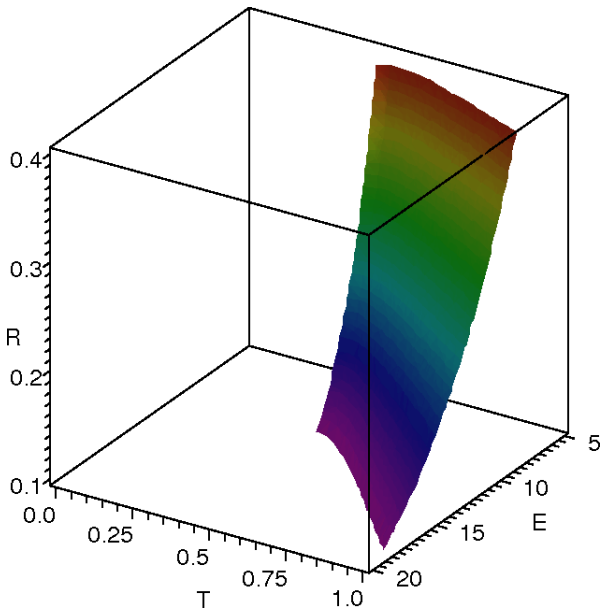}
\caption{Thermodynamic hypersurface of a heterotic string gas at fixed $S$ in a universe with three non- compact spatial dimensions. The variables $E$, $R$ and $T$ are expressed in string units $2 \pi^{2} \sqrt{\alpha'}$. This hypersurface exists only for temperature, $T > T_{H}=(1+\sqrt{2})^{-1}$.} \label{fig:1}
\end{figure} 

At this point we should address the question how realistic this non-compact string gas system is. We have assumed from the outset that winding modes are absent in non-compact directions. In fact the winding modes can get excited at high energies giving rise to large-radius corrections that in turn can modify $\Omega(E, V)$ \cite{Deo3}. The temperature of the system shows an unphysical behavior too. It approaches the Hagedorn temperature from above as the energy density $\rho \rightarrow \infty$. The Hamilton-Jacobi equation given in Eq. (\ref{eq:hamilton-j}) becomes ill defined once the temperature of the system falls below the Hagedorn temperature. In that case the thermodynamic hypersurface $F(q, p)=0$ and thus the family of string gas does not exist.
\subsection{String Gas in a Compact and Expanding Universe} 
It is interesting to know whether a thermodynamic hypersurface of co-dimension one exists when the string gas lives in a compact universe. String gas thermodynamics during the quasi-equilibrium expansion of a string universe is addressed in \cite{Deo3}. All the $D$ spatial dimensions are compact with toroidal geometry and $d$ of them expand in circles of radii $R_{i}$, $i=1, \cdots, d$. The remaining $D-d$ spatial dimensions are small and of the order of $\sqrt{\alpha'}$. 

From \cite{Deo3} we have the density of states in the thermodynamic limit (space is compact but large and energy density is finite) 
\beq
\label{eq:compact}
\Omega(E, V) \simeq \beta_{H} e^{\beta_{H}E + a_{0}V}~\Big[1 - \frac{(\beta_{H}E)^{\alpha}}{\alpha!}e^{-B(E-\sigma_{0}V)}\Big],
\eeq
where $a_{0} = \lambda_{0} + a \alpha'^{-d/2}$; $\sigma_{0} = \rho_{0} + 2^{5/2}b\beta_{H}^{-1}\alpha'^{-d/2}$, $a$ and $b$ are positive numbers depending only on $d$; $V = (2 \pi R)^{d}$; $\alpha = (2d-1)$ and $B \equiv \beta_{H}-\beta_{1}$. For the case of heterotic superstring with all $R_{i}$ equal, the singularities associated with the partition function in the complex $\beta$ plane is given by \cite{Deo3, Deo4},
\beq
\beta_{n} =(2 \pi^{2} \alpha')^{1/2}\Big[\sqrt{1-\frac{2 \pi^{2} \alpha' n^{2}}{V^{2/d}}}+\sqrt{2-\frac{2 \pi^{2} \alpha' n^{2}}{V^{2/d}}}\Big],
\eeq
where $n$ is an integer. Also note that $\beta_{1}$ is real and $\beta_{1}<\beta_{0} \equiv \beta_{H}$. Thus we have 
\beq
B = (2 \pi^{2} \alpha')^{1/2}\Big[1+\sqrt{2}-\sqrt{1-\frac{2 \pi^{2} \alpha'}{V^{2/d}}}-\sqrt{2-\frac{2 \pi^{2} \alpha'}{V^{2/d}}}\Big].
\eeq

In Eq. (\ref{eq:compact}) we have assumed that the energy density $E/V > \sigma_{0}$ and $d \geq 3$ such that 
\beq
\delta \Omega_{(1)} \equiv \frac{(\beta_{H}E)^{\alpha}}{\alpha!}e^{-B(E-\sigma_{0}V)} \ll 1.
\eeq 

From Eq. (\ref{eq:compact}) we find that the entropy is given by
\beq
\label{eq:entropy-comp}
S(E, V) \simeq \beta_{H}E + a_{0}V - \delta \Omega_{(1)}
\eeq
up to a constant.

The other two equations of state are
\bea
\label{eq:temp-comp}
\Big(\frac{\partial S}{\partial E}\Big)_{V} &\simeq& \beta_{H} - \Big[(2d-1)E^{-1}-B\Big]\delta \Omega_{(1)},\\
\label{eq:press-comp}
\Big(\frac{\partial S}{\partial V}\Big)_{E} &\simeq& a_{0} + \Big[B'(E - \sigma_{0}V) - \sigma_{0}B\Big]\delta \Omega_{(1)},
\eea
where $B' = \partial B/\partial V$.

Entropy given in Eq. (\ref{eq:entropy-comp}) obeys the non-degeneracy condition
\beq
\left| \begin{array}{cc}
\frac{\partial^{2} S}{\partial E\partial \sigma_{0}}  & \frac{\partial^{2} S}{\partial E\partial a_{0}} \\
\frac{\partial^{2} S}{\partial V\partial \sigma_{0}} & \frac{\partial^{2} S}{\partial V\partial a_{0}}  \end{array} \right| = BV(\alpha E^{-1} - B)\delta \Omega_{(1)} \neq 0,
\eeq
for all finite $V$ and $\alpha/E \neq B$. In that case a family of hypersurfaces exists in the thermodynamic phase space. 

The two parameters $a_{0}$ and $\sigma_{0}$ can be eliminated from the three equations of state to get a hypersurface connecting all the thermodynamic variables
\begin{widetext}
\beq
\label{eq:hyper-comp}
\Big[\frac{1}{T_{H}} - \frac{1}{T}\Big]\Big[\frac{EV}{\alpha - EB}\Big]\Big[\frac{1}{V}+EB'-\Big(\frac{B'}{B}+\frac{1}{V}\Big)\Big\{\textrm{ln}\Big(\Big[\frac{1}{T_{H}} - \frac{1}{T}\Big]\frac{\alpha!}{(\alpha - EB)(\beta_{H})^{\alpha}E^{\alpha-1}}\Big)+EB\Big\}\Big]-\frac{PV}{T}-\beta_{H}E+S = 0.
\eeq
\end{widetext}
It should be noted that the compact string universe exhibits a realistic behavior in temperature as long as the energy density is sufficiently high. In other words, if energy $E$ is held fixed, the radii should be allowed to vary only within the range $(c\sqrt{\alpha'}E)^{-1}<\Big(R/\sqrt{\alpha'}\Big)^{d}<c\sqrt{\alpha'}E$, where $c$ is a dimensionless number of order unity \cite{Deo3}. The total energy becomes insufficient to support momentum (winding) modes once the radii are too small (large).  

For the thermodynamic hypersurface given in Eq. (\ref{eq:hyper-comp}) to be well defined we should have $T < T_{H}$ and $E > \alpha/B$. For the coordinates $S,E,V$ and the conjugate momenta $p_{E}, p_{V}$ the thermodynamic hypersurface takes the form
\begin{widetext}
\beq
\label{eq:hyper-comp1}
\Big[\beta_{H} - p_{E}\Big]\Big[\frac{EV}{\alpha - EB}\Big]\Big[\frac{1}{V}+EB'-\Big(\frac{B'}{B}+\frac{1}{V}\Big)\Big\{\textrm{ln}\Big(\Big[\beta_{H} - p_{E}\Big]\frac{\alpha!}{(\alpha - EB)(\beta_{H})^{\alpha}E^{\alpha-1}}\Big)+EB\Big\}\Big]-V p_{V}-\beta_{H}E+S = 0.
\eeq
\end{widetext}

This is the Hamilton-Jacobi equation describing the family of string gases where each member is identified by the choice of parameters $a_{0}$ and $\sigma_{0}$. Again, $\beta_{H}$ is a universal constant for each family.
\begin{figure}
\includegraphics[height=8cm]{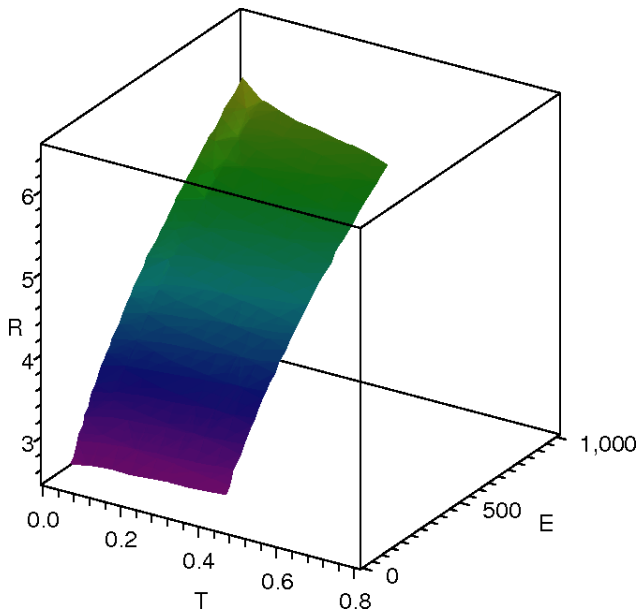}
\caption{
Thermodynamic hypersurface of a heterotic string gas at fixed $S$ in a compact universe with three large space dimensions. The variables $E$, $V$ and $T$ are expressed in string units $2 \pi^{2} \sqrt{\alpha'}$. This hypersurface exists when temperature, $T < T_{H}=(1+\sqrt{2})^{-1}$ and radius $R$ is within the allowed energy range.} \label{fig:2}
\end{figure}

The dynamics of thermodynamics is again on the hypersurface $F(q, p)=0$. The characteristic curves can be obtained from Eq. (\ref{eq:curves}). The parameters, $a_{0}$ and $\sigma_{0}$, characterizing each family members are given by the initial conditions of this dynamics. The `time' variable of this dynamics should be related to the entropy of the system \cite{Rajeev2}.

The above hypersurface is given for the case of large $E$ and large radius $R \gg \sqrt{\alpha'}$. It is interesting to know what happens in the dual theory, that is, when $R \ll \sqrt{\alpha'}$. The spectrum of the theory exhibits duality and thus the density of states also should exhibit the same \cite{Deo3}. That is,
\beq
\Omega (E, V) = \Omega(E, \tilde{V}),
\eeq
where $\tilde{V}=(2 \pi \tilde{R})^{d}$, $\tilde{R}$ being the dual radius. The right hand side of Eq. (\ref{eq:compact}) is given by the same expression but now $V$ is replaced by $\tilde{V}$. The fundamental equation involves $S(E, \tilde{V})$. The other two equations of state follow from
$(\partial S/\partial E )_{\tilde{V}}$ and $(\partial S/\partial \tilde{V})_{E}$ with $B' = \partial B/\partial {\tilde V}$. It is easy to see that a thermodynamic hypersurface similar to Eq. (\ref{eq:hyper-comp}) exists in the dual case also. 

The shortest route from classical mechanics to quantum mechanics is given by the Hamilton-Jacobi formulation. Similarly, in classical thermodynamics also a Hamilton-Jacobi formulation hints towards the possibility of quantum thermodynamics \cite{Rajeev1}. Upon quantization, the generating function $F$ becomes an operator $\hat{F}$. It is interesting to study the quantum version of the string gas thermodynamics and the nature of quantum thermodynamic fluctuations in the string gas system. If we assume that the very early universe was filled with a hot gas of strings \cite{Brandenberger1}, the imprints of quantum thermodynamic fluctuations in this gas might be encoded in the cosmic microwave background radiation and in the distribution of large scale structure in the universe. It would be interesting to investigate the quantum nature of the string thermodynamic system in this scanerio and thus the possibility of observing its effects.  
\section{Conclusions}
In this paper we have derived the Hamilton-Jacobi equations for the thermodynamics of an ideal gas of strings at high energy densities when the topology of space is both non-compact and compact. It is shown that for these two cases there  exist a family of hypersurfaces for the string gas system, characterized by the vanishing of a function $F$. Each member of the family is represented by the chosen values of the two parameters of the system, which have dimensions of number density and energy density. It is also shown that for the non-compact case the family of hypersurfaces exists only for temperatures above Hagedorn temperature revealing the unphysical nature of the system. When the universe is compact, the family of hypersurfaces is well defined when temperature is below the Hagedorn temperature and the size of the universe is within a given bound determined by the energy content of the universe. It is clearly of interest to study the case of string thermodynamics when other conserved charges are also present. We haven't discussed the dynamics of string gas thermodynamics on the hypersurface $F(q, p)=0$. It would be interesting to proceed in this direction as well.  

Hamilton-Jacobi formulation of classical thermodynamics leads to the route to quantum thermodynamics. It is also interesting to investigate the effects of quantum thermodynamic fluctuations of a system of hot string gas in the context of the cosmology of very early universe and the possibility of observing its imprints on cosmic microwave background radiation and the distribution of large scale structure in the universe.
\section{Acknowledgements}
We thank Robert Brandenberger and Chung-I Tan for useful correspondence and Eric West for a careful reading of the manuscript and helpful comments. A.J. thanks the Department of Physics and Astronomy, University of Rochester for their friendly hospitality. S.G.R.'s work is supported in part by a grant from the U.S. Department of Energy under the contract number DE-FG02-91ER40685. A.J.'s work is supported in part by the U.S. Department of Energy grant under the contract number DE-FG02-85ER40231.

\end{document}